\documentclass[11pt,twoside,a4paper]{article}

\usepackage{bm}
\usepackage{ulem}
\usepackage{amsmath}
\usepackage{amsfonts}
\usepackage{relsize}
\usepackage{appendix}
\usepackage{amsthm}
\usepackage{color}
\usepackage{afterpage}
\usepackage{graphicx}
\usepackage{booktabs}
\usepackage{hyperref}
\usepackage[backend=biber, style=numeric]{biblatex}
\begin{document}

\title{Nonparametric covariate hypothesis tests for the cure rate in mixture cure models}

\author{Ana L\'opez-Cheda*
  \and
 M. Amalia J\'acome
  \and
  Ingrid Van Keilegom
  \and
  Ricardo Cao
}

\newcommand{\Addresses}{{
 \footnotesize
  \bigskip
  \footnotesize
  Ana L\'opez-Cheda (Corresponding author), \textsc{Department of Mathematics}, \orgname{University of A Coru\~na}, \orgaddress{\state{Campus de Elvi\~na s/n, 15071 A Coru\~na}, \country{Spain}}\par\nopagebreak
  
  \medskip

  M. Amalia J\'acome, \textsc{Department of Mathematics}, \orgname{University of A Coru\~na}, \orgaddress{\state{Campus de Elvi\~na s/n, 15071 A Coru\~na}\par\nopagebreak}

  \medskip

  Ingrid Van Keilegom, \textsc{\orgdiv{ORSTAT}, \orgname{KU Leuven}, \orgaddress{\state{Naamsestraat 69, 3000 Leuven}, \country{Belgium}}}\par\nopagebreak

}}

\maketitle



\abstract{In lifetime data, like cancer studies, there may be long term survivors, which lead to heavy censoring at the end of the follow-up period. Since a standard survival model is not appropriate to handle these data, a cure model is needed. In the literature, covariate hypothesis tests for cure models are limited to parametric and semiparametric methods. We fill this important gap by proposing a nonparametric covariate hypothesis test for the probability of cure in mixture cure models. A bootstrap method is proposed to approximate the null distribution of the test statistic. The procedure can be applied to any type of covariate, and could be extended to the multivariate setting. Its efficiency is evaluated in a Monte Carlo simulation study. Finally, the method is applied to a colorectal cancer dataset.} 


\maketitle

\footnotetext{\textbf{Abbreviations:} CHUAC, Complexo Hospitalario Universitario de A Coru\~{n}a; CM, Cram\'er-von Mises; CRC, colorectal cancer; K, Kolmogorov-Smirnov; KM, Kaplan-Meier; NW, Nadaraya-Watson}

\section{Cure models}\label{sec:cure}
Classical methods to analyze lifetime data assume that all subjects would experience the failure if there is no censoring and they are followed for long enough. They do not consider the possibility of a group of nonsusceptible individuals that will not develop such event and can be considered as cured. However, there is an increasingly large number of situations where there are individuals who can be deemed to be immune to the event of interest. One well-known example of long-term survivors is cancer studies.

In the literature, the most popular cure model is the mixture cure model (recent reviews of cure models can be found in Peng and Taylor \cite{PengTaylor2014}, and Amico and Van Keilegom \cite{AmicoVanKeilegom2018}, among others). Mixture cure models, proposed by Boag \cite{Boag1949}, split the population into two groups: the cured, who will never experience the event of interest, and the susceptible group. These models allow to estimate the probability of being cured and the survival function of the uncured population, or latency, depending on a set of covariates. The main advantage of this model is that it allows covariates to have different influence on cured and uncured patients. A detailed review of this model is provided by Maller and Zhou.\cite{MallerZhou1996} The estimation of cure models has been extensively studied using parametric and semiparametric methods.\cite{Farewell1986,PengDear2000,SyTaylor2000,ZhangPeng2009,PengTaylor2011,Wangetal2012,PatileaVanKeilegom2017} A nonparametric estimator for the probability of cure \cite{XuPeng2014,LopezChedaetal2017a} and for the latency function \cite{LopezChedaetal2017a,LopezChedaetal2017b} in the mixture cure model was recently introduced and some asymptotic properties further studied. This enables the mixture cure model with covariates to be addressed in a completely nonparametric way.

It is always of interest to test if a covariate has some influence on the cure rate or on the survival time of the susceptible patients. M\"uller and Van Keilegom \cite{MullerVanKeilegom2018} propose a test statistic to assess whether the cure rate, as a function of the covariates, satisfies a certain parametric model. However, to the best of our knowledge, no hypothesis tests for covariate effects in mixture cure models has been proposed yet in a completely nonparametric way. To fill this important gap, a covariate hypothesis test for the probability of cure is presented in this paper. The method, evaluated in a Monte Carlo simulation study, is based on a covariate hypothesis test for nonparametric regression.\cite{DelgadoGonzalezManteiga2001}

The methodology is applied to a real dataset related to colorectal cancer patients. We worked with a dataset related to 414 colorectal cancer (CRC) patients from the University Hospital of A Coru\~na (CHUAC), Spain. The variable of interest is the time (in months) since diagnosis until death from cancer. An individual is considered long-term survivor or cured if he or she will not die because of colorectal cancer. Censoring is caused by ``cure'', death due to any other cause different to colorectal cancer, dropout, or end of the study.

The information provided is, along with the observed lifetime and the censoring indicator, the location (colon $n_1=111$, rectum $n_2=303$), the age (23-102 years) and the stage (1-4), which is the main determinant in prognosis of these patients. The stage has $3$ components: T (related to the size of the tumor and whether it has invaded nearby tissue), N (which measures the lymph nodes that are involved) and M (referring to the presence of metastasis). These components are combined so that we can classify each patient in a unique stage from $1$ to $4$. About $50\%$ of the observations are censored, with the percentage of censoring depending on the stage. The number of patients in Stage $1$ is $62$ ($70.97\%$ censored, aged 23-84), in Stage $2$ is $167$ ($55.09\%$ censored, aged 36-102), in Stage $3$ is $133$ ($39.85\%$ censored, aged 30-88) and $52$ in Stage $4$ ($30.77\%$ censored, aged 43-88).

Cure models should be applied when there is a strong rationale for the existence of cured subjects. Colorectal cancer is one of the leading causes of cancer mortality and morbidity worldwide, accounting for $9.4\%$ of all cancer cases and 1 million new cases annually.\cite{BoyleLevin2008} Death rates from CRC have declined progressively over the last decades, due to improvements such as earlier diagnosis and better treatments, most notably surgical techniques, such as laparoscopic surgery and total mesorectal excision.\cite{Mitryetal2005} This increased the rates of long-term survivors, conventionally been defined by those with at least 5-year survival times after cancer diagnosis.\cite{ACS2017} The long-term survival rate for colorectal cancer patients is, for example, $90.3\%$ when colorectal cancers are detected at a localized stage.\cite{DeSantisetal2014} This shows feasible the possibility of cure in the CRC patients of the analyzed dataset.

\begin{figure}[t]
\centerline{\includegraphics[width=0.45\textwidth]{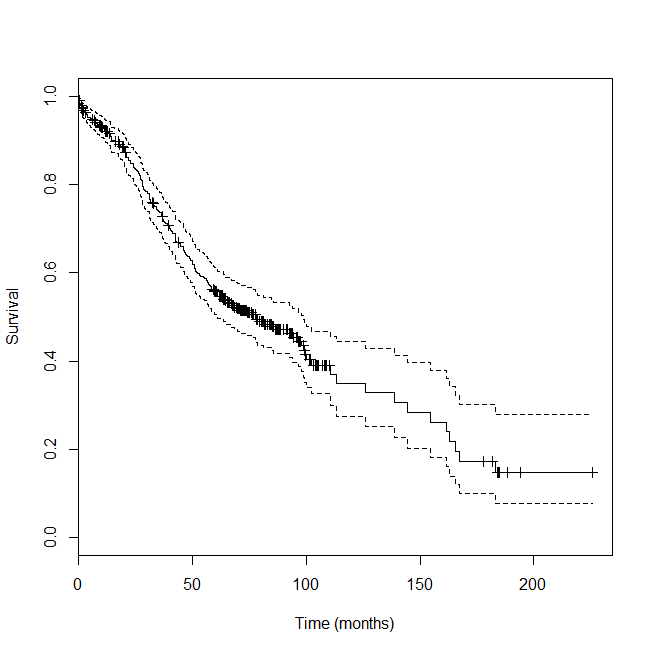}}
\caption{Survival function estimation for the colorectal cancer dataset computed with the Kaplan-Meier estimator. The black crosses correspond to censored observations.\label{fig:art_Nocure}}
\end{figure}

Cure models usually require not only biological evidence for the possibility of cure, but also large sample sizes and a reasonably long follow-up time.\cite{Farewell1986} This is particularly important if censoring is heavy, since too
much censoring or insufficient follow-up time can lead to overestimated cure rates.\cite{LaskaMeisner1992} There are several ways to guess if the follow-up period (almost 19 years in the dataset) is long enough. A characteristic of a cure model is that the limit of the survival function is non-zero as time tends to infinity. Then if the Kaplan-Meier (KM) plot suggests a non-zero asymptote, then a cure model may be
appropriate and useful. Figure \ref{fig:art_Nocure} shows the KM estimate for the survival function for the colorectal cancer dataset. We can appreciate that the survival curve has a plateau at the end of the study. This non-zero asymptote could be taken as an estimator of the cure rate, that is, the proportion of patients who will not die from colorectal cancer, so they can be considered as \textquotedblleft cured\textquotedblright.

Maller and Zhou \cite{MallerZhou1992} proposed a simple nonparametric test to devise whether the follow-up time is enough. The procedure is based on the length of the 
interval at the right tail where the KM estimator of the survival function is constant. A long and stable plateau with heavy censoring at the tail of the KM curve is taken as an evidence that the
follow-up period has been sufficient. The results of the test \cite{MallerZhou1992} suggest that there is a significant evidence for the existence of a cure rate ($p=0.0008589$).

The rest of the article is organized as follows. In Section \ref{sec:not_incidence} we introduce the notation and we give a detailed description of the nonparametric mixture cure model.\cite{XuPeng2014,LopezChedaetal2017a,LopezChedaetal2017b} In Section \ref{sec:sign_tests3cases} we focus on the hypothesis tests for the probability of cure. According to the number of covariates, we address three situations: (a) in Section \ref{sec:Case1} we study if the probability of cure depends on a one-dimensional covariate $Z$ (Case 1); (b) in Section \ref{sec:Case2}, we assume that the probability of cure depends on a one-dimensional covariate $X$, and we study if it also depends on a $m$-dimensional covariate $\bm{Z}$ (Case 2); and (c) Section  \ref{sec:Case3} addresses the general case of $(\bm{X}, \bm{Z})$, where $\bm{X}$ is $\mathbb{R}^q$-valued and $\bm{Z}$ is $\mathbb{R}^m$-valued (Case 3). The good performance of the test was assessed in a simulation study. We also applied the proposed methodology to the real dataset related to colorectal cancer patients from CHUAC (Complexo Hospitalario Universitario de A Coru\~{n}a), Spain.

\section{Nonparametric mixture cure models}
\label{sec:not_incidence}
Throughout this paper we assume that individuals are subject to random right censoring, and that the censoring time, $C$, and the time to occurrence of the event, $Y$, are conditionally independent given a set of covariates, say $\bm{X}$. The conditional distribution function of $Y$ is $F(t|\bm{x})=P(Y\leq t | \bm{X}=\bm{x})$, and the corresponding survival function is $S(t|\bm{x})=1-F(t|\bm{x})$. The observed time is $T=\min (Y,C)$, and $\delta = I(Y \leq C)$ is the uncensoring indicator. Moreover, the conditional distribution functions of $C$ and $T$ are $G$ and $H$, respectively. Let us denote by $\nu$ the cure indicator, with $\nu=0$ if the individual is susceptible to the event, and $\nu =1$ otherwise (it is cured). Note that if $\nu=1$, it is assumed that $Y = \infty$. The probability of not being cured (incidence) is $p(\bm{x}) = P(\nu =0 |\bm{X}=\bm{x})$, and the conditional survival function of the uncured group, also called latency, is  $S_0(t|\bm{x})=P(Y>t | \nu=0, \bm{X} = \bm{x})$. Then, the mixture cure model becomes:
\begin{displaymath}
S(t|\bm{x})=1 - p(\bm{x}) + p(\bm{x}) S_0 (t|\bm{x}).
\end{displaymath}
Without loss of generality, let $X$ be a univariate continuous covariate with density function $f_X$. The observations will be $\{(X_i, T_i, \delta_i), i=1, \dots, n\}$, i.i.d. copies of the random vector $(X, T, \delta)$.

Xu and Peng \cite{XuPeng2014} introduced the following kernel type cure rate estimator:
\begin{equation}
\label{ec:p_estimation} 1-\hat{p_h}(x)= \prod_{i=1}^{n} \left( 1 -
\frac{\delta_{[i]} B_{h[i]}(x)}{\sum_{r=i}^n B_{h[r]}(x) } \right )
= \hat S_h(T_{\max}^1|x),
\end{equation}
where $\hat S_h(t|x)$ is the conditional product-limit estimator \cite{Beran1981},
\begin{equation*}    
B_{h[i]}(x) = \frac{K_h (x - X_{[i]})}{ \sum_{j=1}^n K_h(x - X_{[j]})}
\end{equation*}
are the Nadaraya-Watson (NW) weights with $K_h(\cdot)=\frac{1}{h}K \left (\frac{ \cdot}{h} \right )$ the rescaled kernel with bandwidth $h \rightarrow 0$ and $T_{\max}^1=\max_{i:\delta_i=1}(T_i)$ is the largest uncensored failure time. Here $T_{(1)}  \leq \ldots \leq T_{(n)}$ are the ordered observed times $T_i$'s, and $\delta_{[i]}$ and $X_{[i]}$ are the corresponding uncensoring indicator and covariate concomitants. The estimator (\ref{ec:p_estimation}) has been proved to be the local maximum likelihood estimator of the cure rate \cite{LopezChedaetal2017a}, consistent and asymptotically normal.\cite{XuPeng2014} Furthermore, López-Cheda et al \cite{LopezChedaetal2017a} obtained and i.i.d. representation, found an asymptotic expression of the mean squared error, and proposed a bootstrap selector for the smoothing parameter $h$.

Mixture cure models might have identifiability issues due to problems associated with the tail of the distribution $F$. If the follow-up is not long enough, events could plausibly occur after the longest observed time, making very difficult to distinguish between cured subjects and long-term uncured subjects. Some conditions on the latency component, $S_{0}$, should be imposed to make the mixture cure model identifiable when the cure rate $1-p\left(x\right) $ is completely unspecified.\cite{Lietal2001,HaninHuang2014} Some of these conditions cover the conditional latency function to be fully parametrically specified, or the dependence of $S_{0}$ on the covariate through a proportional hazards form \cite{Wangetal2012} or an accelerated failure time structure.\cite{ChenDu2018} When the latency function is completely unspecified, identifiability is guaranteed if $S_{0}$ is a proper survival function, that is, if $S_{0}\left( \tau ^{-}|x\right) =0\,\ $ for all $x$, where $\tau >0$ is the length of the observation period, including the possibility of $\tau =\infty$.\cite{HaninHuang2014} According to Hanin and Huang \cite{HaninHuang2014}, the assumption of a proper latency function $S_{0}$ is similar to the zero-tail constraint:\cite{PengDear2000,LopezChedaetal2017a,Taylor1995}
\begin{equation}
\tau _{0}(x)\leq \tau _{G}\left( x\right) \text{ for all }x,  \label{tau0}
\end{equation}
where $\tau_{0}\left(x\right) \,$\ and $\tau _{G}\left( x\right) $ are the right extremes of the supports of $S_{0}\left( t|x\right) $ and $G\left( t|x\right)$, respectively. Assumption (\ref{tau0}) states that there is zero probability of a susceptible individual surviving longer than the largest possible censoring time. Identifiability of the model when applying the
proposed test is entailed with assumption (\ref{tau0}), that is, essentially assuming that all the observations after the largest failure time are cured.

To be confident that condition (\ref{tau0}) is true, the length of follow-up should be chosen with considerable care. A nonparametric test for assumption (\ref{tau0}) was proposed by Maller and Zhou.\cite{MallerZhou1992} The idea is based on the difference between the largest observed time, $T_{\left( n\right) }$, and the largest uncensored time, $T_{\max }^{1}$, that is, the interval at the right tail of the distribution where the KM estimator of the survival function, $S$, has got a long stable plateau. A large interval with heavy censoring is considered an evidence that the follow-up period has been sufficiently long for the assumption (\ref{tau0}) to be true. 

\section{Hypothesis tests for the cure rate}
\label{sec:sign_tests3cases}

Testing the effect of a covariate is of primary importance in regression analysis, because the number of potential covariates to be included in the model can be extremely large. In particular, in mixture cure models, variable selection is of great interest, since the covariates having an effect on the survival of the uncured patients are not ne\-ce\-ssa\-rily the same as those impacting the probability of cure. We propose a nonparametric covariate hypothesis test for the cure rate based on a test for selecting explanatory variables in nonparametric regression without censoring.\cite{DelgadoGonzalezManteiga2001} The main advantage over other smoothed tests is that it only requires a smooth nonparametric estimator of the regression function depending on the explanatory variables present under the null hypothesis. This feature is computationally convenient and partially solves the problem of the ``curse of dimensionality'' when selecting regressors in a nonparametric context.

Let us denote by $\bm{W}=\left( \bm{X},\bm{Z}\right) =\left(X_{1},\dots,X_{q},Z_{1},\dots,Z_{m}\right) $ the explanatory co\-va\-riates. We would like to test if the cure probability, as a function of the covariate vector $\bm{W}$, only depends on $\bm{X}$ but not on $\bm{Z}$:
\begin{equation}
\label{eq:testCase1}
H_{0}:E\left( \nu|\bm{X},\bm{Z}\right) \equiv 1-p\left(\bm{X}\right) \text{ vs. }  H_{1}:E\left( \nu|\bm{X},\bm{Z}\right) \equiv 1-p\left(\bm{X},\bm{Z}\right),
\end{equation}
where the function $p(\bm{x},\bm{z})$ depends not only on $\bm{x}$ but also on $\bm{z}$.

Different cases are considered in this paper, depending on the dimension of the covariates: (a) Case 1, where $\bm{W}=Z$ is univariate (Section \ref{sec:Case1}); (b) Case 2, where $\bm{W}=(X, \bm{Z})$, with a one-dimensional covariate $X$ and an $m$-dimensional covariate $\bm{Z}$ (Section \ref{sec:Case2}), and (c) the general Case 3, with $\bm{W}=(\bm{X},\bm{Z})$ where $\bm{X}$ is $\mathbb{R}^q$-valued and $\bm{Z}$ is $\mathbb{R}^m$-valued, that can be generalized from Case 2 (see Section \ref{sec:Case3}).

The main challenge of testing (\ref{eq:testCase1}) is that the response variable (the cure indicator, $\nu$), is only partially observed due to the censoring. The uncensored observations are known to be uncured ($\nu=0$), but it is unknown if a censored individual will be eventually cured or not ($\nu$ is missing). The novelty of the proposed test is that this inconvenience is overcome expressing the regression function of the unobservable (and inestimable) response, $\nu$, as a regression function with response $\eta$, which is not observable but estimable. This implies that the test is carried out with the variables $(\bm{W},\hat \eta)$, that is, in a context without censoring.

 Let us define the variable $\eta$, which is a conditional proxy for $\nu$,  as follows:
\begin{equation}
\label{ec:eta}
\eta=\frac{\nu(1-I(\delta=0,T\leq \tau ))}{1- G\left( \tau |\bm{W}\right) },
\end{equation}
where $\tau$ is an unknown time beyond which a subject can be considered cured. It is easy to check that $E(\eta|\bm{W})=E(\nu|\bm{W})$ if the distribution of $(C | \bm{W}, \nu=0)$ equals that of $(C | \bm{W}, \nu=1)$. Specifically,
\begin{eqnarray*}
E(\eta|\bm{W}) &=& E(\eta|\bm{W}, \nu=0) P(\nu=0|\bm{W}) + E(\eta|\bm{W}, \nu=1) P(\nu=1|\bm{W}).
\end{eqnarray*}
Since $\nu=0$ implies $\eta=0$, then $E(\eta|\bm{W}, \nu=0)=0$ and $E(\eta|\bm{W})$ reduces to
\begin{eqnarray}
\label{ec:E_nu1X}
E(\eta|\bm{W}) &=& \frac{E(\nu (1 - I(\delta=0, T \leq \tau)) |\bm{W}, \nu=1)}{1 - G(\tau|\bm{W})} P(\nu=1|\bm{W}).
\end{eqnarray}
Note that $\nu=1$ implies $\delta=0$, hence $T=C$ and the numerator in (\ref{ec:E_nu1X}) is,
\begin{eqnarray*}
&& E(\nu (1 - I(\delta=0, T \leq \tau)) |\bm{W}, \nu=1) \\
&=& E(1 - I(C \leq \tau) |\bm{W}, \nu=1) = E(I(C>\tau) |\bm{W}, \nu=1)  \\
&=& P(C > \tau|\bm{W}, \nu=1) = 1 - G(\tau| \bm{W},\nu=1).
\end{eqnarray*}
Therefore, if $C$ and $\nu$ are independent conditionally on $\bm{W}$, then
\begin{eqnarray*}
&& 1 - G(\tau|\bm{W},\nu=1) \\
&=& P(C > \tau| \bm{W}, \nu=1) [P( \nu=1 |\bm{W}) + P(\nu=0 | \bm{W})] \\
&=& P(C > \tau|\bm{W}, \nu=1) P(\nu=1 |\bm{W})  + P(C > \tau|\bm{W}, \nu=0) P(\nu=0 |\bm{W}) \\
&=& P(C > \tau| \bm{W}) = 1 - G(\tau|\bm{W}).
\end{eqnarray*}
As a consequence, $E(\eta|\bm{W})$ in (\ref{ec:E_nu1X}) is
\begin{eqnarray*}
E(\eta|\bm{W}) &=& \frac{1-G(\tau|\bm{W}, \nu=1)}{1-G(\tau|\bm{W})} P(\nu=1|\bm{W}) = P(\nu=1|\bm{W}) = E(\nu|\bm{W}).
\end{eqnarray*}

Note that $\{\eta_i, i=1,\dots,n\}$ are not observable because $\tau$ and the conditional distribution function $G(t|\bm{W})$ are not known. Therefore $\tau$ and $G(t|\bm{W})$ have to be estimated to obtain an estimation of $\{\eta_i, i=1,\dots,n\}$. Several estimators of the conditional distribution $G(t|\bm{W})$ can be considered according to the dimension and type of the covariate vector $\bm{W}$. As an alternative to, among others, the popular Cox proportional hazards model, we propose a completely nonparametric approach. Specifically, in Case 1 ($\bm{W}=Z$), if $Z$ is continuous, $G(t|z)$ can be estimated using the conditional product-limit estimator \cite{Beran1981} with a cross-validation (CV) bandwidth selector.\cite{Geerdensetal2018} Otherwise, when $Z$ is discrete or qualitative with values $\{z_1,\ldots,z_k\}$, the stratified KM estimator can be used for every subsample $Z=z_j,j=1,\ldots,k$. In Cases 2 and 3, the conditional distribution $G(t|\bm{w})$ can be estimated nonparametrically according to the type of the covariates $\bm{W}$ following the general ideas in Racine and Li.\cite{RacineLi2004} For example, in the simplest scenario of a bivariate covariate $\bm{W}=(X,Z)$, if both covariates are continuous, $G(t|\bm{w})$ can be estimated with the generalization of the conditional product-limit estimator \cite{LiangdeUnaAlvarez2012} with a cross-validation (CV) bandwidth selector \cite{Geerdensetal2018}, whereas if both $(X,Z)$ are discrete or qualitative, with the stratified KM estimator using the corresponding subsamples. 

The estimation of $\tau$, a cure threshold beyond which a censored observation can be assumed as cured, would seem an ill-posed problem since whether an individual is cured or not is not always observable. However, note that under condition (\ref{tau0}), the largest observed survival time, $T^1_{\max}$, converges in probability to $\tau_0$.\cite{XuPeng2014} Therefore, condition (\ref{tau0}) guarantees asymptotically that all subjects censored after $T^1_{\max}$ can be assumed to be cured. As a consequence, we suggest to estimate in practice $\tau$ as the largest uncensored failure time, $\hat \tau = T^1_{\max}$. 

As a result, the estimation of the values $\{\eta_i,\;i=1, \dots, n\}$ are the following: if $\delta_i=1$, or if $\delta_i =0$ and $T_{i}\leq \hat \tau$, then $\hat \eta_{i}=0$; otherwise $\hat \eta_{i}=1/ (1- \hat G(\hat \tau|\bm{W}_i))$. The test can also be applied even when there is no cure, that is, when $E(\nu|\bm{W}) = E(\eta | \bm{W}) = 0$. In that case, $\hat \tau = T^1_{\max}$ will be close to the largest observed time $T_{(n)}$, and the estimates of $\eta$ in (\ref{ec:eta}) will be mostly zero. This would yield values of the test statistics, to be introduced in next sections, close to zero, suggesting to keep the null hypothesis that the cure rate does not depend on the covariate $\bm{W}$.

\section{Case 1}
\label{sec:Case1}

In this case we study if the cure rate, as a function of $\bm{W}=Z$, is a constant value versus if it depends on the covariate $Z$:
\begin{equation*}
H_{0}: E\left( \nu|Z\right) =1-p \text{ \ constant \ vs } H_1:E\left( \nu|Z\right)=1-p(Z),
\end{equation*}
where $p(z)$ is not a constant function. Using the observations $\{(Z_i,\hat \eta_i),i=1,\ldots,n\}$, the test we propose is based on the following process:
\begin{equation}
\label{ec:sign_tests_T_Case1}
U_{n}(z)=\frac{1}{n}\sum_{i=1}^{n} \left ( \hat{\eta}_{i}- \left ( \frac{1}{n}\sum_{j=1}^{n}\hat{\eta}_{j} \right ) \right )I\left( Z_{i}\leq z\right ),
\end{equation}
which is a mean of the difference between the estimates of $\eta$ and the conditional mean of $\eta$ under the null hypothesis. Possible test statistics are the Cram\'er-von Mises (CM) test, $CM_n=\sum_{i=1}^nU_n^2(Z_i)$, or the Kolmogorov-Smirnov (K) test, $K_n=\max_{i=1,\ldots,n}|n^{1/2}U_n(Z_i)|$. The null distribution of the test statistic is approximated by the boots\-trap procedure. The bootstrap resampling plan used is similar to the one for bandwidth selection in nonparametric incidence and latency estimation but mimicking the null hypothesis.\cite{LopezChedaetal2017a,LopezChedaetal2017b}

The steps of the method are described below. The bootstrap resampling plan mimics $H_0$, since in Step 2 the cured observations are generated with constant probability $1-\hat p$. We propose to estimate the cure rate as $1-\hat p = \hat S_n^{KM}(\infty)$, the KM estimator of the survival function $S(t)=P(Y> t)$ evaluated at the largest uncensored observation. Note that to generate an uncured observation, the conditional distributions $F_0(t|z)$ and $G(t|z)$ in Steps 2.1 and 2.2 have to be estimated. When $Z$ is continuous, suitable estimators are the nonparametric latency estimator\cite{LopezChedaetal2017a} and the conditional PL estimator \cite{Beran1981}, respectively. Otherwise, these functions can be estimated using the corresponding stratified KM estimators. The method proceeds as follows:
\begin{enumerate}
\item[1.] {For $i= 1, 2, \ldots , n$, obtain $Z_i^*$ in $\{Z_1,\ldots,Z_n\}$ by random resampling with replacement.}
\item[2.] {Let $1-\hat p$ be an estimation of the cure probability. For $i= 1, 2, \ldots , n$:
\begin{enumerate}
\item[2.1] {Obtain a bootstrap cured observation $Y_i^*=\infty$ with probability $1-\hat p$, and draw $Y_i^*$ from a nonparametric estimator of the conditional distribution $\hat F_0(t|Z_i^*)= 1- \hat S_0(t|Z_i^*)$ otherwise.}
\item[2.2] {Draw $C_i^*$ from a nonparametric estimator of the conditional distribution $G(t|Z_i^*)$.}
\item[2.3] {Compute $T_i^*=\min(Y_i^*,C_i^*)$ and $\delta_i^*=I(Y_i^* \leq C_i^*)$.}
\end{enumerate}
}
\item[3.] {With the bootstrap resample, $\{(Z_i^*, T_i^*, \delta_i^*),i=1,\ldots,n\}$, compute $\hat \eta_i^*$ in (\ref{ec:eta}), obtain the bootstrap version of $U_n$ in (\ref{ec:sign_tests_T_Case1}), and the corresponding bootstrap version of the Cram\'er-von Mises and Kol\-mo\-go\-rov-Smirnov statistics, $CM_n^*$ and $K_n^*$.
}
\item[4.] {Repeat $B$ times Steps 1-3 in order to generate $B$ values of $CM_n^*$ and $K_n^*$. Define the critical values $d^*_{CM}$ and $d^*_K$ as the values which are in position $\lceil (1-\alpha) B \rceil$ in the corresponding sorted vector.
}
\item[5.] {Compare the value of the statistic, $CM_n$ (respectively, $K_n$),
obtained with the original sample with $d^*_{CM}$ (respectively, $d^*_K$),
and reject the null hypothesis if $CM_n>d^*_{CM}$ (respectively, $K_n>d^*_K$).
In addition, the $p$-value can be calculated as the proportion of resamples for which the bootstrap statistic, $CM_n^*$ (respectively $K_n^*$) is larger than the value of  the statistic with the original sample, $CM_n$ (respectively $K_n$).
}
\end{enumerate}

In the case of $Z$ a non-ordinal qualitative covariate with values $\{z_1,\ldots,z_k\}$, there is no natural way to order the values of $Z$ from lowest to highest. This makes it impossible to compute the indicator function in the test statistic (\ref{ec:sign_tests_T_Case1}). We propose to consider all the possible $k!$ permutations of the values of $Z$ and compute $U_n(z)$ (and the corresponding $CM_n$ and $K_n$ statistics) for each ``ordered'' permutation. Finally, the maximum of the $k!$ values $CM_n$ and $K_n$ is computed and compared with the critical point obtained by the bootstrap likewise.

A different approach consists in working with $k-1$ dummy variables. The main benefit of this method would be that the value of the statistic is computed $k-1$ times, whereas with the previous method, the statistic should be computed $k!$ times. Therefore, when the number $k$ of levels is high, this approach is considerably less computationally expensive. However, the clear advantage of the first approach is that the categorical covariate is tested as a whole, regardless the number of levels. For the simulation study in Section \ref{sec:sim_Case1_Zqualit} the first approach was considered.

\subsection{Simulation study}
\label{sec:sim_sign_Case1}

The purpose of the simulation study was to assess the practical behavior of the proposed test in different scenarios according to the covariate vector $\bm{W}$. We considered  $\kappa=2000$ trials of sample sizes $n=50$, $100$, $200$ and $500$. A total of $B=2000$ bootstrap resamples were drawn. The nominal significance level was $\alpha =0.05$ along the scenarios. All the results were obtained using a script implemented in \texttt{R} language.\cite{R2018}

For Case 1 with $\bm{W}=Z$, we investigated the finite sample behavior of the test for $Z$ continuous, discrete and nominal. The censoring variable, $C$, had conditional distribution $G(t|z) \sim Exp(\lambda(z))$, with $\lambda(z)=0.6/(2+(z-20)/40)$, and the survival function of the uncured individuals was
\begin{equation}
\label{eq:S0_case1}
S_0(t|z) = \frac{ \exp(-\alpha(z) t) - \exp(-\alpha(z) \tau_0) }{1 -
\exp(-\alpha(z) \tau_0)}  I(t \leq \tau_0), 
\end{equation}
where $\tau_0 = 4.605$ and $\alpha \left( z\right) =\exp \left((z+20) /40 \right)$.

\subsubsection{$Z$ continuous}
\label{sec:sim_Case1_Zcont}
Let $Z$ be a continuous random variable with distribution $U(-20,20)$. Under the null hypothesis, $H_0: E(\nu|Z)=1-p$, we considered four different scenarios: $p=0.5,\;0.6,\;0.7,\;0.8$. We also considered the case of no cure ($p=1$). Under the alternative hypothesis, the cure probability was
\begin{equation}
\label{ec:M1_Case1}
1 - p(z) = 1 - \frac{ \exp ( 0.476 + 0.358 z ) }{ 1 + \exp (0.476 + 0.358 z)}.
\end{equation}
The average percentage of censored data was $54.65\%$ and of cured data was $46.66\%$.

To estimate the conditional distribution function $G(t|z)$, required in the estimation of $\eta$ in (\ref{ec:eta}) and in Step 2.2 of the bootstrap procedure, and $F_0(t|z)$, needed in Step 2.1 of the bootstrap, we used the conditional product-limit \cite{Beran1981} and the conditional latency \cite{LopezChedaetal2017a} estimators, respectively. The bandwidth was selected, with the CV procedure \cite{Geerdensetal2018}, from a grid of 10 equispaced bandwidths $h_j=D_j n^{-1/5}$, from $D_1=4$ to $D_{10}=60$.

The results are given in Table \ref{tab:Case1ZCont}. It is noteworthy that, under $H_0$, the size of the test is quite close to the nominal level $\alpha=0.05$ for all the values of $p$, even for large censoring rates ($p=0.5$) and small sample sizes ($n=50$). As expected, when there is no cure, the test is conservative regardless the sample size, keeping the null hypothesis that the cure probability does not depend on $Z$. Furthermore, under $H_1$, the power of the test is very close to 1 for all the sample sizes, even for $n=50$.

\begin{center}
\begin{table}[t]
\caption{Results of the test for Case 1 with $\bm{W}=Z$ continuous with distribution $U(-20,20)$, under the null and the alternative hypotheses, respectively. The case without cure has also been considered.\label{tab:Case1ZCont}}
\centering
\resizebox{\columnwidth}{!}{%
\begin{tabular}{ccccccccccc|cc}
\hline
\multicolumn{11}{c|}{$H_{0}$} & \multicolumn{2}{|c}{$H_{1}$} \\ \hline
& \multicolumn{2}{c}{$p=0.5$} & \multicolumn{2}{c}{$p=0.6$} & 
\multicolumn{2}{c}{$p=0.7$} & \multicolumn{2}{c}{$p=0.8$} & 
\multicolumn{2}{c|}{$p=1$} & \multicolumn{2}{|c}{} \\ 
& \multicolumn{2}{c}{%
\begin{tabular}{c}
60.4\% cens \\ 
50\% cure%
\end{tabular}%
} & \multicolumn{2}{c}{%
\begin{tabular}{c}
52.5\% cens \\ 
40\% cure%
\end{tabular}%
} & \multicolumn{2}{c}{%
\begin{tabular}{c}
44.6\% cens \\ 
30\% cure%
\end{tabular}%
} & \multicolumn{2}{c}{%
\begin{tabular}{c}
36.6\% cens \\ 
20\% cure%
\end{tabular}%
} & \multicolumn{2}{c|}{%
\begin{tabular}{c}
21\% cens \\ 
Without cure%
\end{tabular}%
} & \multicolumn{2}{|c}{%
\begin{tabular}{c}
54.6\% cens \\ 
46.7\% cure%
\end{tabular}%
} \\ \hline
$n$ & CM & K & CM & K & CM & K & CM & K & CM & K & CM & K \\ \hline
$50$ & 0.046 & 0.055 & 0.045 & 0.051 & 0.045 & 0.050 & 0.039 & 0.042 & 0.021
& 0.018 & 0.983 & 0.978 \\ 
$100$ & 0.043 & 0.050 & 0.051 & 0.057 & 0.042 & 0.045 & 0.044 & 0.045 & 0.015
& 0.011 & 0.999 & 0.999 \\ 
$200$ & 0.053 & 0.055 & 0.047 & 0.046 & 0.037 & 0.039 & 0.046 & 0.040 & 0.021
& 0.013 & 1 & 1 \\ 
$500$ & 0.065 & 0.053 & 0.056 & 0.054 & 0.056 & 0.053 & 0.048 & 0.041 & 0.024
& 0.014 & 1 & 1 \\ \hline
\end{tabular}
}
\end{table}
\end{center}

\subsubsection{$Z$ discrete}
\label{sec:sim_Case1_Zdisc}
We considered a discrete covariate $Z$ with $3$ ordered values $z_1<z_2<z_3$. The uncure probabilities $p(z_i)$ were obtained evaluating in $z_i$ the function $p$  in (\ref{ec:M1_Case1}). For each scenario, we chose the values $\{z_j,j=1,2,3\}$ such that $p(z_1)=p(z_2)=p(z_3)\in \{0.5,0.6,0.7,0.8\}$ under $H_0$, including the no cure case  $p(z_1)=p(z_2)=p(z_3)=1$. Under $H_1$, we considered the cases $p(z_1)\in \{0.1, 0.3\}$, $p(z_2)=0.5$ and $p(z_3)\in \{0.7, 0.9\}$. We simulated two situations according to the probability mass function of $Z$ given by $\Pi_z(z_i)=P(Z=z_i)$: $\Pi_z =(1/3,1/3,1/3)$ and $\Pi_z = (1/5,1/5,3/5)$.

The results in Table \ref{tab:Case1:H0_Zdisc} under the null hypothesis suggest that the sample size $n=50$ seems to be quite small in general, specially for low cure rates. However, the rejection rate under $H_0$ increases up to the nominal level $\alpha=0.05$ as the sample size gets larger. Under the alternative hypothesis, the power of the test is quite acceptable even for small sample sizes, being very close to $1$  when the values of the cure rate are more distant from the null hypothesis, that is, for $(p(z_1),p(z_2),p(z_3))=(0.1,0.5,0.9)$. Finally, when there is no cure,  the rejection rates are much smaller than the nominal value (lower than 0.01 for any sample size $n$).

\begin{center}
\begin{table}[t]
\centering
\caption{Size (top) and power (bottom) of the test for Case 1 with $Z$ discrete, under the null and the alternative hypotheses, respectively.\label{tab:Case1:H0_Zdisc}}
\resizebox{\columnwidth}{!}{%
\begin{tabular}{ccccccccccc}
\hline
& \multicolumn{10}{c}{$%
\begin{array}{c}
\text{ } \\ 
H_{0}:E\left( \nu |Z=z_{i}\right) =1-p,\ i=1,2,3 \\ 
\text{ }%
\end{array}%
$} \\ \hline
& \multicolumn{2}{c}{$p=0.5$} & \multicolumn{2}{c}{$p=0.6$} & 
\multicolumn{2}{c}{$p=0.7$} & \multicolumn{2}{c}{$p=0.8$} & 
\multicolumn{2}{c}{$p=1$} \\ 
& \multicolumn{2}{c}{%
\begin{tabular}{c}
60.8\% cens \\ 
50\% cure%
\end{tabular}%
} & \multicolumn{2}{c}{%
\begin{tabular}{c}
52.7\% cens \\ 
40\% cure%
\end{tabular}%
} & \multicolumn{2}{c}{%
\begin{tabular}{c}
44.5\% cens \\ 
30\% cure%
\end{tabular}%
} & \multicolumn{2}{c}{%
\begin{tabular}{c}
36.1\% cens \\ 
20\% cure%
\end{tabular}%
} & \multicolumn{2}{c}{%
\begin{tabular}{c}
14.1\% cens \\ 
Without cure \ \ 
\end{tabular}%
} \\ \hline
$n$ & CM & K & CM & K & CM & K & CM & K & CM & K \\ \hline
$50$ & 0.038 & 0.029 & 0.041 & 0.043 & 0.037 & 0.039 & 0.032 & 0.034 & 0.009
& 0.008 \\ 
$100$ & 0.039 & 0.039 & 0.046 & 0.040 & 0.048 & 0.037 & 0.032 & 0.030 & 0.009
& 0.007 \\ 
$200$ & 0.051 & 0.051 & 0.040 & 0.041 & 0.043 & 0.045 & 0.038 & 0.039 & 0.005
& 0.006 \\ 
$500$ & 0.043 & 0.040 & 0.041 & 0.043 & 0.051 & 0.050 & 0.050 & 0.047 & 0.004
& 0.004 \\ \hline
\bigskip  & \multicolumn{10}{c}{$%
\begin{array}{c}
\\ 
H_{1}:E\left( \nu |Z=z_{i}\right) =1-p_{i},\ i=1,2,3 \\ 
\end{array}%
$} \\ \hline
&  & \multicolumn{4}{c}{$%
\begin{array}{c}
\text{ \ } \\ 
\text{ \ }%
\end{array}%
\left( p\left( z_{1}\right) ,p\left( z_{2}\right) ,p\left( z_{3}\right)
\right) =\left( 0.3,0.5,0.7\right) 
\begin{array}{c}
\text{ \ } \\ 
\text{ \ }%
\end{array}%
$} & \multicolumn{4}{c}{$%
\begin{array}{c}
\text{ \ } \\ 
\text{ \ }%
\end{array}%
\left( p\left( z_{1}\right) ,p\left( z_{2}\right) ,p\left( z_{3}\right)
\right) =\left( 0.1,0.5,0.9\right) 
\begin{array}{c}
\text{ \ } \\ 
\text{ \ }%
\end{array}%
$} &  \\ \cline{2-11}
&  & \multicolumn{2}{c}{$\Pi _{z}=\left( 1/3,1/3,1/3\right) $} & 
\multicolumn{2}{c}{$\Pi _{z}=\left( 1/5,1/5,3/5\right) $} & 
\multicolumn{2}{c}{$\Pi _{z}=\left( 1/3,1/3,1/3\right) $} & 
\multicolumn{2}{c}{$\Pi _{z}=\left( 1/5,1/5,3/5\right) $} &  \\ 
&  & \multicolumn{2}{c}{%
\begin{tabular}{c}
60\% cens \\ 
50\% cure%
\end{tabular}%
} & \multicolumn{2}{c}{%
\begin{tabular}{c}
52.2\% cens \\ 
42\% cure%
\end{tabular}%
} & \multicolumn{2}{c}{%
\begin{tabular}{c}
58.8\% cens \\ 
50\% cure%
\end{tabular}%
} & \multicolumn{2}{c}{%
\begin{tabular}{c}
44.3\% cens \\ 
34\% cure%
\end{tabular}%
} &  \\ \hline
$n$ &  & CM & K & CM & K & CM & K & CM & K &  \\ \hline
$50$ &  & 0.386 & 0.348 & 0.303 & 0.305 & 0.925 & 0.884 & 0.895 & 0.892 & 
\\ 
$100$ &  & 0.680 & 0.635 & 0.614 & 0.604 & 0.991 & 0.987 & 0.986 & 0.986 & 
\\ 
$200$ &  & 0.920 & 0.885 & 0.842 & 0.823 & 0.999 & 0.999 & 0.998 & 0.997 & 
\\ 
$500$ &  & 0.999 & 0.998 & 0.996 & 0.994 & 1 & 1 & 0.999 & 0.999 &  \\ \hline
\end{tabular}
}
\end{table}
\end{center}

\subsubsection{$Z$ qualitative}
\label{sec:sim_Case1_Zqualit}
A qualitative covariate $Z$ with three possible values $\{b_1,b_2,b_3\}$ was considered. For each scenario, $\{b_1,b_2,b_3\}$ were linked to the numerical values $\{z_1,z_2,z_3\}$ such that the values $p(b_1)$, $p(b_2)$ and $p(b_3)$ and the latency functions $S_0(t|b_1)$, $S_0(t|b_2)$ and $S_0(t|b_3)$ were given by $p(z)$ in (\ref{ec:M1_Case1}) and $S_0(t|z)$ in (\ref{eq:S0_case1}) evaluated at $\{z_1,z_2,z_3\}$, respectively. The conditional distribution of the censoring variable was $C|Z=b_1 \sim Exp(0.6)$, $C|Z=b_2 \sim Exp(0.45)$ and $C|Z=b_3 \sim Exp(0.3)$. 

Under $H_0:p(b_1)=p(b_2)=p(b_3)=p$, we used $p\in \{0.5,0.6,0.7,0.8\}$, along with the case of no cure ($p=1$). Under the alternative hypothesis, two scenarios were considered, $(p(b_1), p(b_2),p(b_3))=(0.3,0.5,0.7)$ and $(0.1,0.5,0.9)$. Each scenario was simulated with two possible probability mass functions for $Z$, $(1/3,1/3,1/3)$ and $(1/5,1/5,3/5)$. 

The results, given in Table \ref{tab:Case1:H0_Zqualit}, are very similar to those in the $Z$ discrete case (Table \ref{tab:Case1:H0_Zdisc}). The sample size $n=50$ seems to be small to achieve the nominal level $\alpha=0.05$, specially for low cure rates. However, the rejection rate under $H_0$ with moderate and large sample sizes increases significantly, reaching the nominal value 0.05, faster for larger cure rates. Regarding the alternative hypothesis, as expected, the power of the test is higher for large sample sizes and when $H_1$ is easier to detect, that is, in the most extreme case $(p(b_1), p(b_2),p(b_3))=(0.1,0.5,0.9)$.  As in the previous cases, with no cures the rejection rates of the test are very low.

\begin{center}
\begin{table}[t]
\caption{Size (top) and power (bottom) of the test for Case 1 with $Z$ qualitative with values $\{b_1,b_2,b_3\}$ and probability mass function $\Pi_z=\left ( \Pi_{z}(b_1), \Pi_{z}(b_2),\Pi_z(b_3) \right )$ under the null and the alternative hypotheses, respectively. The results without cure are also given.\label{tab:Case1:H0_Zqualit}}
\centering
\resizebox{\columnwidth}{!}{%
\begin{tabular}{ccccccccccc}
\hline
& \multicolumn{10}{c}{$%
\begin{array}{c}
\text{ } \\ 
H_{0}:E\left( \nu |Z=b_{i}\right) =1-p,\ i=1,2,3 \\ 
\text{ }%
\end{array}%
$} \\ \hline
& \multicolumn{2}{c}{$p=0.5$} & \multicolumn{2}{c}{$p=0.6$} & 
\multicolumn{2}{c}{$p=0.7$} & \multicolumn{2}{c}{$p=0.8$} & 
\multicolumn{2}{c}{$p=1$} \\ 
& \multicolumn{2}{c}{%
\begin{tabular}{c}
68\% cens \\ 
50\% cure%
\end{tabular}%
} & \multicolumn{2}{c}{%
\begin{tabular}{c}
52.7\% cens \\ 
40\% cure%
\end{tabular}%
} & \multicolumn{2}{c}{%
\begin{tabular}{c}
44.5\% cens \\ 
30\% cure%
\end{tabular}%
} & \multicolumn{2}{c}{%
\begin{tabular}{c}
36.1\% cens \\ 
20\% cure%
\end{tabular}%
} & \multicolumn{2}{c}{%
\begin{tabular}{c}
14.1\% cens \\ 
Without cure \ \ 
\end{tabular}%
} \\ \hline
$n$ & CM & K & CM & K & CM & K & CM & K & CM & K \\ \hline
$50$ & 0.036 & 0.039 & 0.033 & 0.037 & 0.039 & 0.042 & 0.024 & 0.027 & 0.008
& 0.009 \\ 
$100$ & 0.043 & 0.043 & 0.041 & 0.040 & 0.041 & 0.039 & 0.028 & 0.029 & 0.006
& 0.006 \\ 
$200$ & 0.052 & 0.048 & 0.039 & 0.041 & 0.041 & 0.043 & 0.037 & 0.037 & 0.010
& 0.010 \\ 
$500$ & 0.042 & 0.041 & 0.043 & 0.046 & 0.039 & 0.041 & 0.048 & 0.045 & 0.004
& 0.004 \\ \hline
\bigskip  & \multicolumn{10}{c}{$%
\begin{array}{c}
\\ 
H_{1}:E\left( \nu |Z=b_{i}\right) =1-p_{i},\ i=1,2,3 \\ 
\end{array}%
$} \\ \hline
&  & \multicolumn{4}{c}{$%
\begin{array}{c}
\text{ \ } \\ 
\text{ \ }%
\end{array}%
\left( p\left( b_{1}\right) ,p\left( b_{2}\right) ,p\left( b_{3}\right)
\right) =\left( 0.3,0.5,0.7\right) 
\begin{array}{c}
\text{ \ } \\ 
\text{ \ }%
\end{array}%
$} & \multicolumn{4}{c}{$%
\begin{array}{c}
\text{ \ } \\ 
\text{ \ }%
\end{array}%
\left( p\left( b_{1}\right) ,p\left( b_{2}\right) ,p\left( b_{3}\right)
\right) =\left( 0.1,0.5,0.9\right) 
\begin{array}{c}
\text{ \ } \\ 
\text{ \ }%
\end{array}%
$} &  \\ \cline{2-11}
&  & \multicolumn{2}{c}{$\Pi _{z}=\left( 1/3,1/3,1/3\right) $} & 
\multicolumn{2}{c}{$\Pi _{z}=\left( 1/5,1/5,3/5\right) $} & 
\multicolumn{2}{c}{$\Pi _{z}=\left( 1/3,1/3,1/3\right) $} & 
\multicolumn{2}{c}{$\Pi _{z}=\left( 1/5,1/5,3/5\right) $} &  \\ 
&  & \multicolumn{2}{c}{%
\begin{tabular}{c}
60.1\% cens \\ 
50\% cure%
\end{tabular}%
} & \multicolumn{2}{c}{%
\begin{tabular}{c}
52.2\% cens \\ 
42\% cure%
\end{tabular}%
} & \multicolumn{2}{c}{%
\begin{tabular}{c}
58.8\% cens \\ 
50\% cure%
\end{tabular}%
} & \multicolumn{2}{c}{%
\begin{tabular}{c}
44.3\% cens \\ 
34\% cure%
\end{tabular}%
} &  \\ \hline
$n$ &  & CM & K & CM & K & CM & K & CM & K &  \\ \hline
$50$ &  & 0.294 & 0.303 & 0.297 & 0.291 & 0.863 & 0.848 & 0.899 & 0.894 & 
\\ 
$100$ &  & 0.575 & 0.579 & 0.589 & 0.579 & 0.990 & 0.989 & 0.987 & 0.989 & 
\\ 
$200$ &  & 0.871 & 0.861 & 0.818 & 0.813 & 1 & 1 & 0.998 & 0.997 &  \\ 
$500$ &  & 0.998 & 0.998 & 0.994 & 0.992 & 1 & 1 & 1 & 1 &  \\ \hline
\end{tabular}
}
\end{table}
\end{center}

\subsection{Application to the CRC dataset}
\label{sec:real_data_Case1_sign_tests}

We firstly started studying if the tumor location ($Z$) had any effect on the cure rate. Since the result of the test was not significant ($p_{CM}=0.180$, $p_{K}=0.434$), for the rest of the analyses we worked with all the individuals regardless the location of the tumor, both colon and rectum. Next we studied, separately, the effect of the covariates age ($Z_1$) and stage ($Z_2$) on the probability of cure. Age at diagnosis and tumor stage are known to strongly influence colorectal cancer treatment regimen and five-year survival.\cite{Vercellietal2006,Guyotetal2005} However, the effect of the age and stage on the probability of cure is rarely analyzed, few studies of cure for colorectal cancer patients have focused on the estimation of cure by age and stage at diagnosis \cite{Shacketal2012}, but not on the statistical significance of those covariates on the probability of cure. Conlon et al \cite{Conlonetal2014} propose a multi-state Markov model with an incorporated cured fraction to assess how specific covariates influence the cure rate. They state that the covariate age does not have any influence, unlike the covariate stage. Furthermore, the studies of cure for colorectal cancer patients usually categorize the age into intervals, not treating it as a continuous covariate. So the purpose was to test whether the age or the stage have a significant effect on the cure rate, using the nonparametric hypothesis test.
\begin{figure}[t]
\begin{center}
\includegraphics[width=0.45\textwidth]{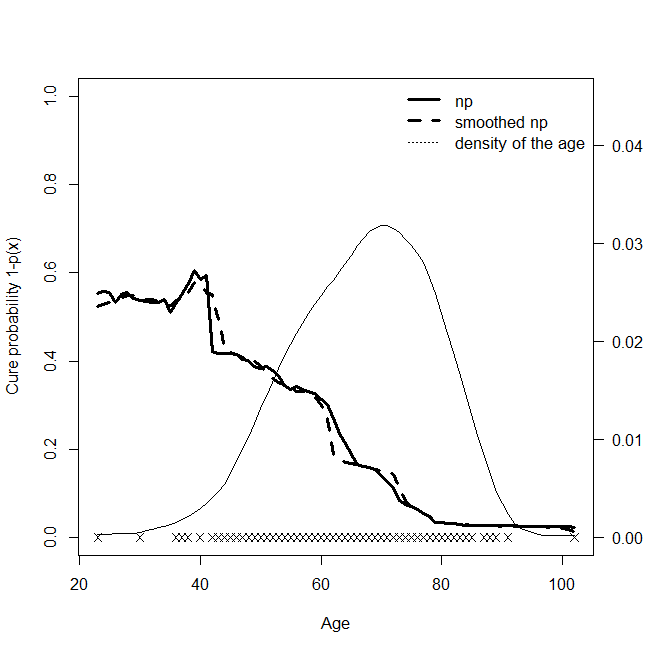} 
\includegraphics[width=0.45\textwidth]{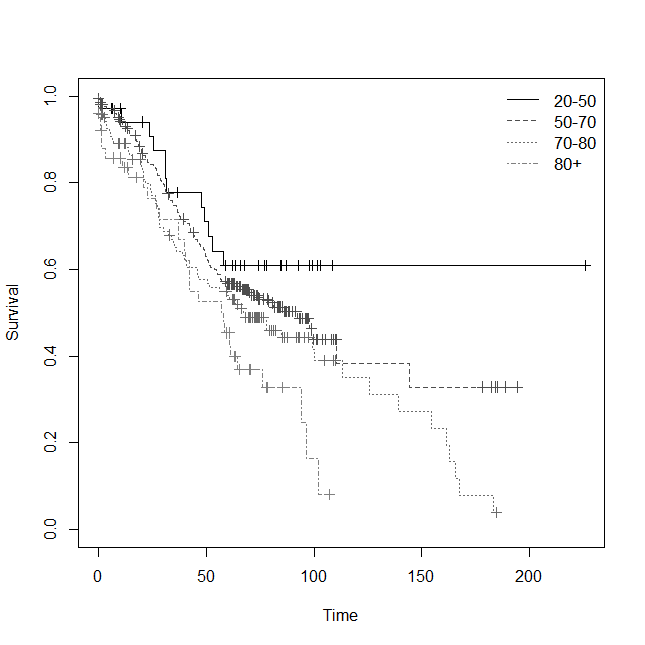}
\end{center}
\caption{Left panel: Nonparametric estimation of the cure probability depending on the age computed with the bootstrap bandwidth (solid line) and with a smoothed bootstrap bandwidth (dashed line). The thin solid line represents the Parzen-Rosenblatt kernel density estimation of the covariate age, using Sheather and Jones’ plug-in bandwidth. Right panel: Estimated KM survival curves for age groups.} \label{incidence}
\end{figure}

\begin{figure}[t]
\begin{center}
\includegraphics[width=0.45\textwidth]{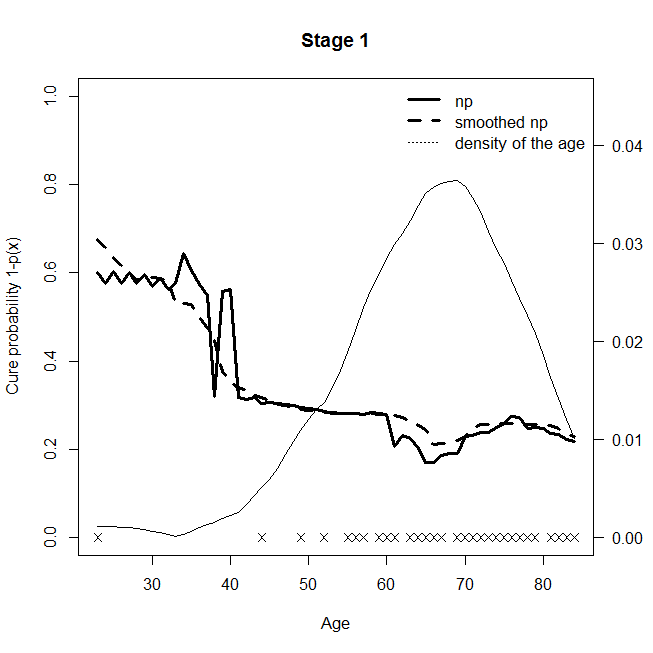} %
\includegraphics[width=0.45\textwidth]{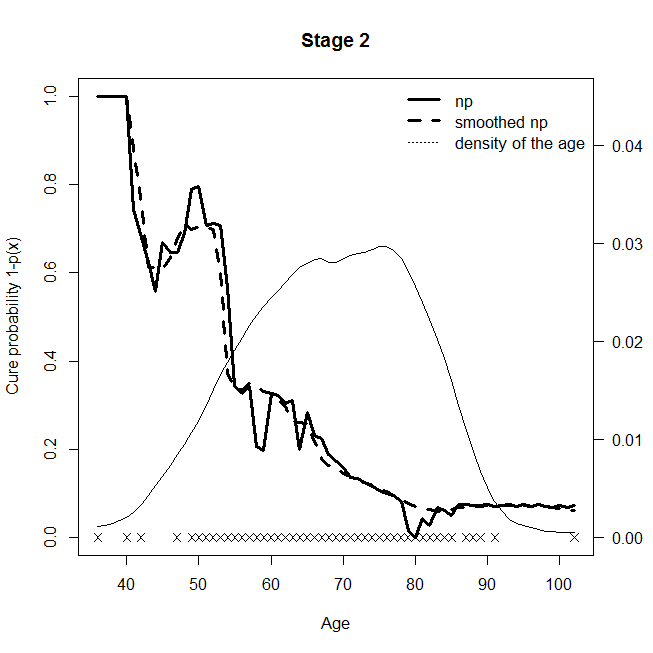} \\[0pt]
\includegraphics[width=0.45\textwidth]{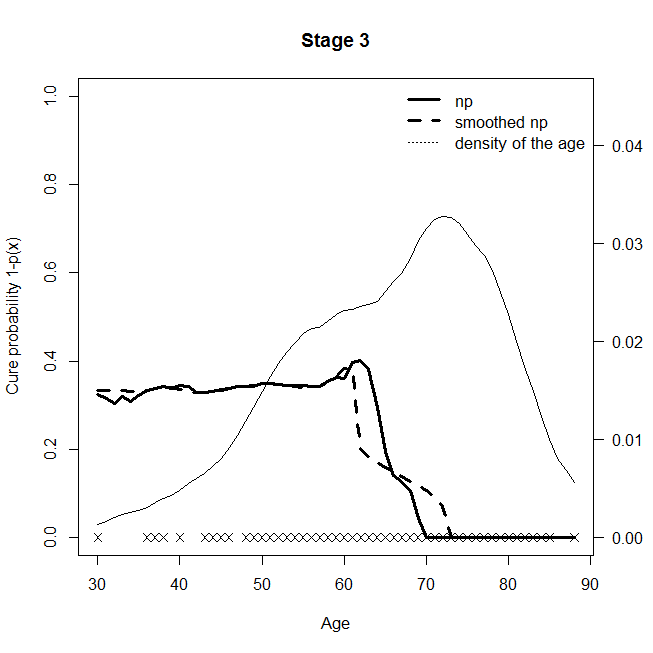} %
\includegraphics[width=0.45\textwidth]{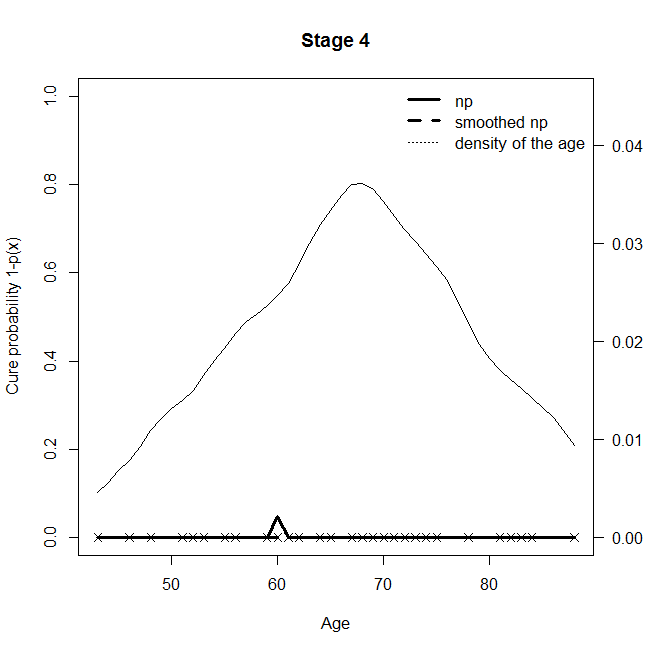} \\[0pt]
\end{center}
\caption{Nonparametric estimation of the cure probability depending on the age for the patients in every stage separately, computed with the boots\-trap bandwidth (solid line) and with a smoothed bootstrap bandwidth (dashed line). The thin solid line represents the Parzen-Rosenblatt kernel density estimation of the covariate age, using Sheather and Jones' plug-in bandwidth.}
\label{incidence_stage1}
\end{figure}

In Figure \ref{incidence} we can appreciate how the nonparametric estimator of the cure rate changes with the age ($Z_1$). In general, the cure probability decreases with increasing age, suggesting that the age may have some influence on the cure rate. Younger patients are likely to tolerate the intensive cancer treatments better than older patients, and therefore they achieve cure with higher probability. On the other hand, elderly patients are substantially less likely to receive surgery and chemotherapy.\cite{Shacketal2012} The test found this effect of the age on the cure probability as significant $(p_{CM}=0.017$, $p_{K}=0.026$).

We also tested the effect of the age on the cure probability within each stage (see Figure \ref{incidence_stage1}). In Stage 1 the cure rate remains almost constant regardless the age (it fluctuates around $25\%$ for most patients), whereas for Stages $2$ and $3$ the age may have some influence since the cure probability decreases as the age increases. Specifically, in Stage 2, this probability decreases with the age, from about $30\%$ in patients with age at diagnosis 50-60 to $7\%$ for patients above 80. Regarding Stage 3, the cure probability is around $35\%$ for individuals younger than 60, whereas for patients above this age that probability decreases dramatically. In Stage 4, the nonparametric estimation of the cure probability is 0. This result suggests that long-term survival for individuals at this stage is uncommon. When the test was applied for each stage separately, the effect of the age on the cure probability was not found significant in any stage ($p_{CM}=0.410$, $p_{K}=0.276$ for Stage 1; $p_{CM}=0.418$, $p_{K}=0.498$ for Stage 2; $p_{CM}=0.186$, $p_{K}=0.166$ for Stage 3; and $p_{CM}=0.767$, $p_{K}=0.767$ for Stage 4).
\begin{figure}[t]
\begin{center}
\includegraphics[width=0.45\textwidth]{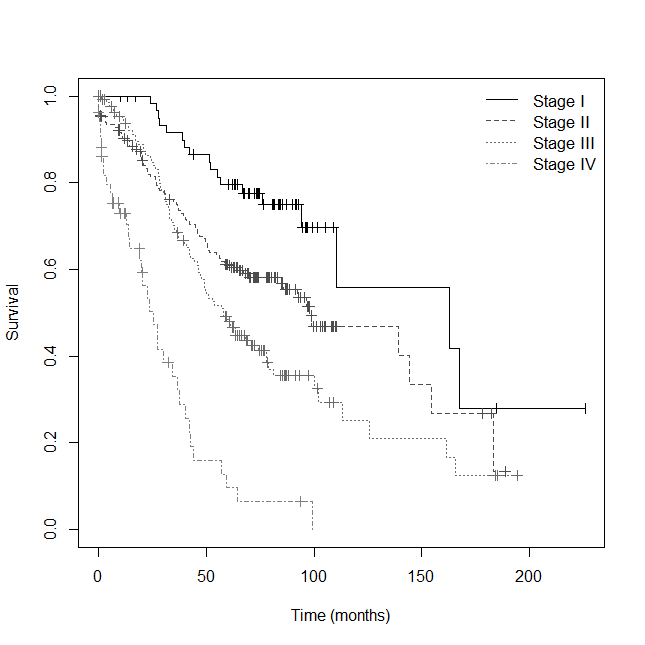} 
\end{center}
\caption{Estimated survival functions according to the stage.} \label{fig:KMstage}
\end{figure}

Regarding the effect of the stage ($Z_2$) on the probability of cure, the more tumors are detected at earlier stage, the more curative resections are possible.\cite{Itoetal2012} The estimated cure probabilities for each stage support that statement, decreasing from $0.28$ in Stage 1 to $0.13$ in Stages 2 and 3, and 0 in Stage 4, see Figure \ref{fig:KMstage}. However, these differences in the cure rate according to the stage were not significant ($p_{CM}=0.475$ and $p_{K}=0.655$).

\section{Case 2}
\label{sec:Case2}
In this case, $\bm{W}=(X, \bm{Z})$ has dimension $m+1$, with a one-dimensional covariate, $X$, and an $m$-dimensional covariate, $\bm{Z}$. We study if the cure probability, as a function of $\bm{W}=(X,\bm{Z})$, only depends on the covariate $X$, that is:
\begin{equation*}
H_{0}: E\left( \nu|X,
\bm{Z}\right) =1-p(X) \text{ \ vs } H_1:E\left( \nu|X,
\bm{Z}\right)=1-p(X, \bm{Z}),
\end{equation*}
where $p(x,\bm{z})$ depends on $\bm{z}$. To do this, we use the observations $\{(X_i, \bm{Z}_i,\hat \eta_i),i=1,\ldots,n\}$. Following \cite{DelgadoGonzalezManteiga2001}, the statistic is defined as:
\begin{equation}
\label{ec:sign_tests_T_Case2}
U_{n}(x,\bm{z})=\frac{1}{n}\sum_{i=1}^{n} \hat f_{X}(X_i) \left
( \hat{\eta}_{i}- \hat m(X_i) \right ) I\left( (X_i,\bm{Z}_i) \leq
(x,\bm{z})  \right),
\end{equation}
where $\hat f_{X}(x)$ is a nonparametric estimator of the density function of the covariate $X$, $\hat m(x)$ is a nonparametric estimator of the regression function $m(x)=E \left (\hat \eta |X=x \right )$, and $\leq$ stands for component-wise inequality. Note that the process $U_{n}$ in (\ref{ec:sign_tests_T_Case2}) is a weighted mean of the difference between the $\hat{\eta}_i$ and their conditional mean under the null hypothesis. Similarly to Case 1, we consider the Cram\'er-von Mises, $CM_{n}=\sum_{i=1}^nU_{n}^2(\bm{W}_i)$ and the Kol\-mo\-go\-rov-Smirnov, $K_{n}=\max_{i=1,\ldots,n}|n^{1/2}U_{n}(\bm{W}_i)|$ statistics.

When the covariate $X$ is continuous, the density of $X$ can be nonparametrically estimated with the Parzen-Rosenblatt estimator $\hat f_{X,h}(x)$  which depends on a bandwidth $h$, and the regression function $m(x)$ with the NW kernel estimator $\hat m_h(x)$ computed with the same bandwidth $h$. As a consequence, a bandwidth $h$ needs to be selected for the computation of the test statistic  in (\ref{ec:sign_tests_T_Case2}), say $U_{n,h}$. There are two main approaches for bandwidth selection in smoothing-based hypothesis tests: power maximization under the alternative hypothesis \cite{KulasekeraWang1997}; and mini\-mi\-za\-tion of $p$-values.\cite{MartinezCamblor2010,MartinezCamblordeUnaAlvarez2013} The two approaches are strongly related.\cite{deUnaAlvarez2013} 

For a categorical or discrete variable $X$, the estimated density, $\hat f_{X}(X_i)$, and regression function, $\hat m(X_i)$, in (\ref{ec:sign_tests_T_Case2}) are replaced by
\begin{displaymath}
\hat \Pi(X_i)=\frac{1}{n} \sum_{j=1}^n I(X_j=X_i) \text{ and } \hat m(X_i)=\frac{\frac{1}{n} \sum_{j=1}^n I(X_j=X_i) \hat \eta_j}{\hat \Pi(X_i)},
\end{displaymath}
respectively. Therefore, in this case the process $U_n$  in (\ref{ec:sign_tests_T_Case2}) does not depend on any smoothing parameter $h$. 

Similarly as in Case 1, for any qualitative variable in $\bm{W}=(X, \bm{Z})$ with no intrinsic order in their values, the indicator function $I\left (\bm{W}_i \leq \bm{w} \right)$ in (\ref{ec:sign_tests_T_Case2}) is computed for all the possible ``ordered'' permutations of the values of the qualitative covariates.

The distribution of the test under $H_0$ is approximated by the bootstrap, con\-si\-de\-ring the following procedure, parallel to the algorithm presented in Section \ref{sec:Case1}:
\begin{enumerate}
\item[1.] {For $i= 1, 2, \ldots , n$, obtain $(X_i^*,\bm{Z}_i^*)$ from $\{(X_1,\bm{Z}_1),\ldots,(X_n,\bm{Z}_n)\}$ by random resampling with replacement.
}
\item[2.] {For $i= 1, 2, \ldots , n$, compute a nonparametric estimator of the cure probability, $1-\hat p(X_i^*)$. Then:
\begin{enumerate}
\item[2.1] {With probability $1-\hat p(X_i^*)$ generate a bootstrap cured observation $Y_i^*=\infty$. Otherwise, $Y_i^*$ is drawn from a nonparametric estimator of $F_0(t|X_i^*,\bm{Z}_i^*)=1-S_0(t|X_i^*,\bm{Z}_i^*)$.}
\item[2.2] {Draw $C_i^*$ from a nonparametric estimator of the conditional
distribution $G(t|X_i^*,\bm{Z}_i^*)$.}
\item[2.3] {Compute $T_i^*=\min(Y_i^*,C_i^*)$ and $\delta_i^*=I(Y_i^* \leq C_i^*)$.}
\end{enumerate}
}
\item[3.] {With the bootstrap resample, $\{(X_i^*, \bm{Z}_i^*, T_i^*, \delta_i^*),i=1,\ldots,n\}$, compute $\hat \eta_i^*$ in (\ref{ec:eta}), obtain the bootstrap version of $U_{n}$ in (\ref{ec:sign_tests_T_Case2}), and the bootstrap version of the Cram\'er-von Mises and Kol\-mo\-go\-rov-Smirnov statistics, say $CM_{n}^*$ and $K_{n}^*$.
}
\end{enumerate}

Steps 4-5 are the same as those in the bootstrap procedure of Section \ref{sec:Case1}. Note that in order to mimic $H_0$, the bootstrap resamples are generated assuming in Step 2 that the cure rate does not depend on $\bm{Z}$.

As in Case 1, a nonparametric estimator of the cure probability, $1-p(x)$, is needed is Step 2. When $X$ is continuous, we propose to use the nonparametric estimator \cite{XuPeng2014,LopezChedaetal2017a} $1-\hat p_g(x) = \hat S_g(T^1_{\max}|x)$, the conditional PL estimator of the survival function $S(t|x)$ evaluated at the largest uncensored time, with a CV bandwidth selector.\cite{Geerdensetal2018} For a discrete or qualitative variable $X$ with values $\{x_1,\ldots,x_k\}$, the cure rate $1-p(x_j)=E(\hat\eta|X=x_j)$ can be estimated as the sample mean of the $\{\hat\eta_i,i=1,\ldots,n\}$ such that $X_i=x_j$.

Nonparametric estimators of the functions $F_0(t|x, \bm{z})=1-S_0(t|x, \bm{z})$ and $G(t|x, \bm{z})$ are required in Steps 2.1 and 2.2. These estimators depend on the type of the covariates $\bm{W}=(X,\bm{Z})$, and they can be computed following the methodology in Racine and Li.\cite{RacineLi2004} Specifically, the estimator of $G(t|x, \bm{z})$ is the same as the one considered for the estimation of $\eta$ in (\ref{ec:eta}) (see Section \ref{sec:sign_tests3cases}). The conditional distribution $F_0(t|x, \bm{z})=1-S_0(t|x, \bm{z})$ can be estimated with the generalized conditional product-limit estimator.\cite{LiangdeUnaAlvarez2012} All the aforementioned estimators are computed with the corresponding CV bandwidth selector.\cite{Geerdensetal2018}



\subsection{Simulation study}
\label{sec:sim_sign_Case2}
In this case, $\bm{W}=(X,\bm{Z})$ has dimension $m+1$, with a univariate $X$ and a $m$-dimensional covariate $\bm{Z}$. For simplicity, in this simulation study we assumed that $Z$ was also one-dimensional. The test statistic depends on the type of covariates $X$ and $Z$. For the sake of brevity, we will show only the results for the cases when $(X,Z)$ are both continuous (Section \ref{sec:sim_Case2_XcontZcont}) and discrete (\ref{sec:sim_Case2_XdiscZdisc}). Since the $\bm{W}=(X,Z)$ continuous case was a highly time consuming process, only $B=1000$ bootstrap resamples were generated in that case. 

The censoring variable $C$ had conditional distribution $C|X=x,Z=z \sim Exp(\lambda(x,z))$, with $\lambda(x,z)=0.6/(2+(0.5(x+z)-20)/40)$, and the latency was
\begin{equation*}
S_0(t|x,z) = \frac{ \exp(-\alpha(x,z) t) - \exp(-\alpha(x,z) \tau_0) }{1 - \exp(-\alpha(x,z) \tau_0)}  I(t \leq \tau_0), 
\end{equation*}
where $\tau_0 = 4.605$ and $\alpha \left( x,z\right) =\exp \left((z+20) /40 \right) $ under $H_0$ and $\alpha \left( x,z\right) =\exp \left((x+z+20) /40 \right) $ under $H_1$. The incidence was
\begin{equation}
\label{eq:Case2ec_M1}
p(x,z) = \frac{ \exp (0.476 + 0.358 x (1 + \beta_2 z)) }{ 1 + \exp (0.476 + 0.358 x (1+  \beta_2 z)) },
\end{equation}
with $\beta_2=0$ under $H_0$ and $\beta_2=0.225$ under $H_1$.

\subsubsection{$(X,Z)$ continuous}
\label{sec:sim_Case2_XcontZcont}
We considered two continuous covariates $(X,Z)$ with distribution $N(0,5)$. The conditional distribution functions $F_0(t|x,z)$ and $G(t|x,z)$ were estimated with the ge\-ne\-ra\-lized latency \cite{LopezChedaetal2017a} and product-limit \cite{LiangdeUnaAlvarez2012} estimators, respectively. For the bandwidth required by these estimators, we used the CV bandwidth selector \cite{Geerdensetal2018}, using a search grid of equispaced bandwidths $h_j=D_j n^{-1/6}$, from $D_1=3.5$ to $D_{10}=30$. The performance of the test $U_{n,h}$  was assessed in a grid of bandwidths $h=Cn^{-1/3m}$, following Delgado and Gonz\'alez-Manteiga \cite{DelgadoGonzalezManteiga2001}, where $C$= 10, 20, 30, 40, 45, 50, 60, and $m$ was the dimension of the tested covariate $\bm{Z}$ (note that in our case, $m=1$).

The results are given in Table \ref{tab:Case2_Xcont_Zcont_H0}. The performance of the test under $H_0$ is quite acceptable for any sample size if the bandwidth $h$ is suitably chosen, specially as the sample size increases. In that case, the dependence of the results on the bandwidth $h$ seems to weaken. The power of the test under $H_1$ is quite high regardless the bandwidth, increasing, as expected, with the sample size. 

\begin{center}
\begin{table}[t]
\caption{Size (top) and power (bottom) of the test for Case 2 with $X$ and $Z$ continuous with distribution $N(0,5)$, under the null and the alternative hypotheses, respectively.\label{tab:Case2_Xcont_Zcont_H0}}
\centering
\begin{tabular}{cccccc}
\hline
\multicolumn{2}{c}{} & \multicolumn{2}{c}{$H_{0}$} & \multicolumn{2}{c}{$%
H_{1}$} \\ 
&  & \multicolumn{2}{c}{%
\begin{tabular}{c}
52.8\% cens \\ 
42.4\% cure%
\end{tabular}%
} & \multicolumn{2}{c}{%
\begin{tabular}{c}
53.2\% cens \\ 
43.1\% cure%
\end{tabular}%
} \\ \hline
$n$ & $h=Cn^{-1/3}$ & CM & K & CM & K \\ \hline
$%
\begin{array}{c}
50 \\ 
\\ 
\\ 
\\ 
\  \\ 
\  \\ 
\ 
\end{array}%
$ & $%
\begin{array}{c}
2.71 \\ 
5.43 \\ 
8.14 \\ 
10.86 \\ 
12.21 \\ 
13.57 \\ 
16.28%
\end{array}%
$ & $%
\begin{array}{c}
0.013 \\ 
0.015 \\ 
0.018 \\ 
0.032 \\ 
0.039 \\ 
0.041 \\ 
0.048%
\end{array}%
$ & $%
\begin{array}{c}
0.021 \\ 
0.028 \\ 
0.040 \\ 
0.066 \\ 
0.075 \\ 
0.085 \\ 
0.095%
\end{array}%
$ & $%
\begin{array}{c}
0.061 \\ 
0.081 \\ 
0.088 \\ 
0.091 \\ 
0.088 \\ 
0.085 \\ 
0.084%
\end{array}%
$ & $%
\begin{array}{c}
0.069 \\ 
0.093 \\ 
0.097 \\ 
0.110 \\ 
0.118 \\ 
0.112 \\ 
0.108%
\end{array}%
$ \\ 
$%
\begin{array}{c}
100 \\ 
\\ 
\\ 
\\ 
\  \\ 
\  \\ 
\ 
\end{array}%
$ & $%
\begin{array}{c}
2.15 \\ 
4.31 \\ 
6.46 \\ 
8.62 \\ 
9.69 \\ 
10.78 \\ 
12.93%
\end{array}%
$ & $%
\begin{array}{c}
0.024 \\ 
0.030 \\ 
0.036 \\ 
0.044 \\ 
0.047 \\ 
0.049 \\ 
0.057%
\end{array}%
$ & $%
\begin{array}{c}
0.034 \\ 
0.040 \\ 
0.057 \\ 
0.077 \\ 
0.091 \\ 
0.096 \\ 
0.116%
\end{array}%
$ & $%
\begin{array}{c}
0.444 \\ 
0.488 \\ 
0.544 \\ 
0.580 \\ 
0.584 \\ 
0.590 \\ 
0.600%
\end{array}%
$ & $%
\begin{array}{c}
0.375 \\ 
0.432 \\ 
0.443 \\ 
0.450 \\ 
0.450 \\ 
0.450 \\ 
0.437%
\end{array}%
$ \\ 
$%
\begin{array}{c}
200 \\ 
\\ 
\\ 
\\ 
\  \\ 
\  \\ 
\ 
\end{array}%
$ & $%
\begin{array}{c}
1.71 \\ 
3.42 \\ 
5.13 \\ 
6.84 \\ 
7.69 \\ 
8.55 \\ 
10.26%
\end{array}%
$ & $%
\begin{array}{c}
0.036 \\ 
0.034 \\ 
0.036 \\ 
0.032 \\ 
0.036 \\ 
0.040 \\ 
0.039%
\end{array}%
$ & $%
\begin{array}{c}
0.033 \\ 
0.042 \\ 
0.043 \\ 
0.062 \\ 
0.070 \\ 
0.079 \\ 
0.100%
\end{array}%
$ & $%
\begin{array}{c}
0.777 \\ 
0.802 \\ 
0.830 \\ 
0.845 \\ 
0.847 \\ 
0.848 \\ 
0.853%
\end{array}%
$ & $%
\begin{array}{c}
0.712 \\ 
0.739 \\ 
0.748 \\ 
0.735 \\ 
0.722 \\ 
0.706 \\ 
0.685%
\end{array}%
$ \\
$%
\begin{array}{c}
500 \\ 
\\ 
\\ 
\\ 
\  \\ 
\  \\ 
\ 
\end{array}%
$ & $%
\begin{array}{c}
1.26 \\ 
2.52 \\ 
3.78 \\ 
5.04 \\ 
5.67 \\ 
6.30 \\ 
7.56%
\end{array}%
$ & $%
\begin{array}{c}
0.042 \\ 
0.045 \\ 
0.046 \\ 
0.050 \\ 
0.050 \\ 
0.050 \\ 
0.057%
\end{array}%
$ & $%
\begin{array}{c}
0.056 \\ 
0.063 \\ 
0.062 \\ 
0.060 \\ 
0.068 \\ 
0.081 \\ 
0.096%
\end{array}%
$ & $%
\begin{array}{c}
0. 990\\ 
0.989 \\ 
0.994 \\ 
0.995 \\ 
0.995 \\ 
0.996 \\ 
0.997%
\end{array}%
$ & $%
\begin{array}{c}
0.977 \\ 
0.977 \\ 
0.978 \\ 
0.973 \\ 
0.968 \\ 
0.969 \\ 
0.959%
\end{array}%
$ \\ \hline
\end{tabular}
\end{table}
\end{center}

\subsubsection{$(X,Z)$ discrete}
\label{sec:sim_Case2_XdiscZdisc}
The covariates $X$ and $Z$ are discrete variables with values $\{x_1,x_2,x_3\}$ and $\{z_1,z_2,z_3\}$, respectively.  For any scenario, the values of $X$ were $\{x_1,x_2,x_3\}=\{-2.4622, -0.19702, 1.0371\}$. The values of $Z$ were $z_1=z_2=z_3=0.6157$ under $H_0$, and   $\{z_1,z_2,z_3\}=\{-13.123, 0, 4. 9454\}$ under $H_1$. The probabilities $p(x_i,z_j)$, with $i,j=1,2,3$, are given by $p$ in (\ref{eq:Case2ec_M1}) evaluated at $(x_i, z_j)$, see Table \ref{tab:p_xz_H0_1_Xdisc_Zdisc} for details.  We simulated two different situations depending on the corresponding probability mass functions for $X$ and $Z$: in the first one, both are $(1/3,1/3,1/3)$, whereas in the second one, both are $(1/5,1/5,3/5)$.

\begin{center}
\begin{table}[t]
\centering
\caption{Uncure probabilities, $\{p(x_i, z_j),\;i,j = 1, 2, 3\}$ for Case 2 when $X$ and $Z$ are discrete with values $\{x_1,x_2,x_3\}$ and $\{z_1, z_2, z_3\}$, respectively. The values  $p(x_i, z_j)$  are obtained evaluating $p$ in (\ref{eq:Case2ec_M1}) in $(x_i, z_j),\;i, j = 1, 2, 3$.}
\label{tab:p_xz_H0_1_Xdisc_Zdisc}
\begin{tabular}{l|c|c|c}
\toprule
$\mathbf{H_0}$ &  $z_1=0.6157$ & $z_2=0.6157$ & $z_3=0.6157$  \\
\midrule
$x_1=-2.4622$ & $0.40$ & $0.40$ & $0.40$   \\
$x_2=-0.19702$ & $0.60$  & $0.60$ & $0.60$   \\
$x_3=1.0371$ & $0.70$  & $0.70$ & $0.70$ \\
\hline
$\mathbf{H_1}$ &  $z_1=-13.123$ & $z_2=0$ & $z_3=4.9454$ \\
 \hline
$x_1=-2.4622$ & $0.10$ & $0.60$ & $0.80$ \\
$x_2=-0.19702$ & $0.35$  & $0.40$ & $0.42$  \\
$x_3=1.0371$ & $0.56$  & $0.30$ & $0.22$  \\
\bottomrule
\end{tabular}
\end{table}
\end{center}

Table \ref{tab:Case2_H0_Xdisc_Zdisc1} shows the results of the test. Under the null hypothesis, the rejection levels are very close to the nominal $\alpha=0.05$, specially for larger sample sizes. It is important to highlight that the power of the test increases considerably with the sample size. For instance, for CM it increases from $0.076$ with $n$=50, to $0.946$ with $n$=500. 

\begin{center}
\begin{table}[t]
\centering
\caption{Size and power of the test for Case $2$ with $X$ and $Z$ discrete, with values $\{x_1,x_2,x_3\}$ and $\{z_1, z_2, z_3\}$, respectively. The probability mass function of $X$, $\Pi _{x}$, equals that of $Z$, $\Pi _{z}$.\label{tab:Case2_H0_Xdisc_Zdisc1}}
\resizebox{\columnwidth}{!}{%
\begin{tabular}{ccccc|cccc}
\hline
& \multicolumn{4}{c|}{$H_{0}$} & \multicolumn{4}{|c}{$H_{1}$} \\ \hline
& \multicolumn{2}{c}{$\Pi _{z}=\left( 1/3,1/3,1/3\right) $} & 
\multicolumn{2}{c|}{$\Pi _{z}=\left( 1/5,1/5,3/5\right) $} & 
\multicolumn{2}{c}{$\Pi _{z}=\left( 1/3,1/3,1/3\right) $} & 
\multicolumn{2}{c}{$\Pi _{z}=\left( 1/5,1/5,3/5\right) $} \\ 
& \multicolumn{2}{c}{%
\begin{tabular}{c}
55.4\% cens \\ 
42.9\% cure%
\end{tabular}%
} & \multicolumn{2}{c|}{%
\begin{tabular}{c}
52.6\% cens \\ 
37.6\% cure%
\end{tabular}%
} & \multicolumn{2}{|c}{%
\begin{tabular}{c}
54.1\% cens \\ 
44.1\% cure%
\end{tabular}%
} & \multicolumn{2}{c}{%
\begin{tabular}{c}
52.9\% cens \\ 
38\% cure%
\end{tabular}%
} \\ \hline
$n$ & CM & K & CM & K & CM & K & CM & K \\ \hline
$%
\begin{array}{c}
50 \\ 
100\  \\ 
\ 200 \\ 
\ 500%
\end{array}%
$ & $%
\begin{array}{c}
0.029 \\ 
0.048 \\ 
0.041 \\ 
0.044%
\end{array}%
$ & $%
\begin{array}{c}
0.044 \\ 
0.059 \\ 
0.056 \\ 
0.049%
\end{array}%
$ & $%
\begin{array}{c}
0.040 \\ 
0.045 \\ 
0.042 \\ 
0.040%
\end{array}%
$ & $%
\begin{array}{c}
0.066 \\ 
0.051 \\ 
0.050 \\ 
0.043%
\end{array}%
$ & $%
\begin{array}{c}
0.076 \\ 
0.231 \\ 
0.494 \\ 
0.946 %
\end{array}%
$ & $%
\begin{array}{c}
0.092 \\ 
0.212 \\ 
0.377 \\ 
0.854 %
\end{array}%
$ & $%
\begin{array}{c}
0.073 \\ 
0.158 \\ 
0.244 \\ 
0.514 %
\end{array}%
$ & $%
\begin{array}{c}
0.099 \\ 
0.149 \\ 
0.225 \\ 
0.449 %
\end{array}%
$ \\ \hline
\end{tabular}
}
\end{table}
\end{center}


\subsection{Application to the CRC dataset}
\label{sec:real_data_Case2_sign_tests}

Since the probability of cure was found to depend on the age of the patient ($X$), see Section \ref{sec:real_data_Case1_sign_tests}, the test was performed to study if it also depended on the cancer stage ($Z$). Note that since the covariate age ($X$) is continuous, a bandwidth $h$ was required to compute the test statistic in (\ref{ec:sign_tests_T_Case2}). The test was applied using a set of bandwidths $h=Cn^{-1/3}$ with $C=10$, $20$, $40$, $60$, $100$, $125$, $150$, $200$, $250$, $300$, $350$, $400$, $450$, $500$, $550$ and $600$. The results indicate that if the age ($X$) is assumed to affect the cure probability, then the effect of the stage ($Z$) was not statistically significant with any of the values of $h$ considered, from the smallest one $h_1=1.34$ ($p_{CM}=0.394$, $p_{K}=0.504$) to the largest one $h_{16}=80.50$ ($p_{CM}=0.068$, $p_{K}=0.073$).

\section{Extensions of the test}
\label{sec:Case3} 
The test can be generalized to a covariate vector $\bm{W}=(\bm{X},\bm{Z})$, where $\bm{X}$ is $\mathbb{R}^q$-valued and $\bm{Z}$ is $\mathbb{R}^m$-valued, as follows:
\begin{equation}
U_{n}(\bm{w}) = \frac{1}{n} \sum_{i=1}^n  \hat f_{\bm{X}} (\bm{X}_i) (\hat \eta_i -  \hat m (\bm{X}_i)) I(\bm{W}_i \leq \bm{w}),
\label{ec:Uncase3}
\end{equation}
where $\hat f_{\bm{X}}(\bm{x})$ and $ \hat m(\bm{x})$ are multidimensional nonparametric estimators of the density function of $\bm{X}$ and the regression function $m(\bm{x}) = E(\hat \eta | \bm{X}=\bm{x})$, respectively.

Dealing with multivariate covariates $\bm{W}$ complicates the nonparametric estimation of the density function $f_{\bm{X}}(\bm{x})$ and the regression function $m(\bm{x})$ needed for the computation of $U_{n}(\bm{w})$ in (\ref{ec:Uncase3}), the estimation of the conditional censoring distribution $G(t|\bm{w})$ required for the estimation of $\eta_i$ in (\ref{ec:eta}), and the estimation of the latency function $S_0(t|\bm{w})$, essential in Step 2.1 of the bootstrap procedure. Suitable nonparametric estimators are available in the literature \cite{DelgadoGonzalezManteiga2001,RacineLi2004,LiangdeUnaAlvarez2012} that avoid the curse of dimensionality using product kernels. Specifically, if $\bm{X}$ is continuous, the density $f_{\bm{X}}(\bm{x})$ and the regression function $m( \bm{x})$ can be nonparametrically estimated as follows:
\begin{equation*}
\hat f_{\bm{X}h} (\bm{X}_i)= \frac{1}{nh^q} \sum_{j=1}^n \mathbb{K} \left ( \frac{\bm{X}_i - \bm{X}_j}{h} \right ) \text{\;and\;} \hat m_h (\bm{X}_i) = \frac{1}{nh^q}  \frac{1}{\hat f_{\bm{X}h} (\bm{X}_i)} \sum_{j=1}^n \mathbb{K} \left ( \frac{\bm{X}_i - \bm{X}_j}{h} \right ) \hat \eta_j,
\end{equation*}
where $\mathbb{K}$ denotes a kernel function on $\mathbb{R}^q$ such as $\mathbb{K}(\bm{x}) = \prod_{j=1}^q K (x_j)$, and $h \rightarrow 0$ is the bandwidth parameter. This general case is not considered in the simulation study.


\section{Conclusions}
\label{s:discuss}
A nonparametric hypothesis test for the effect of covariates $\bm{W} = (\bm{X},\bm{Z})$ on the probability of cure in mixture cure models is introduced. Specifically, $\bm{X}$ is assumed to affect the cure rate and the influence of $\bm Z$ is tested. The methodology, that can be applied to any type of covariates, enjoys the flexibility of nonparametric hypothesis tests, with the advantage of getting rid of the need for a bandwidth parameter when there is not any continuous variable in $\bm{X}$, including the simplest case of no covariate $\bm{X}$. For more complex scenarios, the choice of a smoothing parameter $h$ is required. Several bandwidth selectors for smoothed tests are proposed in the literature that can also be applied in this context.

The test is based on the estimation of an unobservable variable $\eta$, a variable with the same conditional expectation as $\nu$, through suitable estimation of the conditional distribution of the censoring variable $G(t|\bm{w})$, and a cure threshold $\tau$. To approximate the distribution of the test using the bootstrap, the estimation of the cure rate $p(\bm{x})$ and the latency function $S_0(t|\bm{w})$ is also needed. The results of the simulation study support the use of nonparametric estimators for these functions with a CV bandwidth selector.\cite{Geerdensetal2018} The method is applied to a colorectal cancer dataset, and the results show that the covariate age has a significant influence on the cure probability.


\section*{Acknowledgments}
The first author's research was sponsored by the Beatriz Galindo Junior Spanish Grant (code BEAGAL18/00143) from MICINN (Ministerio de Ciencia, Innovación y Universidades) with reference BGP18/00154, and by the Spanish FPU (Formaci\'on de Profesorado Universitario) Grant FPU13/01371 from MECD (Ministerio de Educaci\'on, Cultura y Deporte). All the authors acknowledge partial support by the MINECO (Mi\-nis\-terio de Econom\'ia y Competitividad) Grant MTM2014-52876-R (EU ERDF support included) and the MICINN (Ministerio de Ciencia, Innovaci\'on y Universidades) Grant MTM2017-82724-R (EU ERDF support included). The first, second and fourth authors acknowledge partial support of Xunta de Galicia (Centro Singular de Investigaci\'on de Galicia accreditation ED431G/01 2016-2019 and Grupos de Referencia Competitiva CN2012/130 and ED431C2016-015) and the European Union (European Regional Development Fund - ERDF). Financial support from the European Research Council (2016-2021, Horizon 2020 / ERC grant agreement No. 694409) for the third author is gratefully acknowledged. The authors are grateful to Dr. S. P\'ertega and Dr. S. Pita, at the University Hospital of A Coru\~na, for providing the colorectal cancer dataset, and to two anonymous reviewers whose suggestions were very helpful to improve this paper.
\vspace*{-8pt}



\subsection*{Conflict of interest}

The authors declare no potential conflict of interests.

%
%


\printbibliography

%

\end{document}